\documentclass{JHEP3}

\usepackage{amssymb}
\usepackage{amsfonts}
\usepackage{amsbsy}
\usepackage{amsmath}
\usepackage{amsthm}
\usepackage{graphicx}
\usepackage{epstopdf}
\usepackage[vcentermath]{youngtab}
\usepackage{multirow}
\usepackage{latexsym}
\usepackage{array}
\usepackage{url}

\usepackage[nosort]{cite}

\newcommand\field[1]{{\ensuremath{\mathbb{{#1}}}}}

\newcommand{\ZZ}{\field{Z}}

\newcommand{\MM}{\mathcal{M}}
\newcommand{\NN}{\mathcal{N}}

\newcommand{\p}{\partial}

\newcommand{\OO}{\mathcal{O}}

\def\ov{\over}

\def\ha{{1\over 2}}

\def\vev#1{\langle#1\rangle}

\def\Tr{{\rm Tr}}
\def\NN{{\cal N}}

\def\Res#1{\mathop{\text{Res}}_{#1}}
\def\eq#1{(\ref{#1})}

\def \ra {\rightarrow}

\def\sig{{\sigma}}

\def\Ga{{\Gamma}}

\def\NN{{\cal N}}

\def\q{{\mathfrak{q}}}

\def\d2x{{\left( {d^2 \ov dx^2} \right)_0}}

\def\e{\epsilon}

\newcommand{\be}{\begin{equation}}
\newcommand{\ee}{\end{equation}}
\newcommand{\bea}{\begin{eqnarray}}
\newcommand{\eea}{\end{eqnarray}}

\newcommand{\bln}{\begin{align}}
\newcommand{\eln}{\end{align}}
\newcommand{\bst}{\begin{split}}
\newcommand{\est}{\end{split}}
\newcommand{\bi}{\begin{itemize}}
\newcommand{\ei}{\end{itemize}}
\newcommand{\ben}{\begin{enumerate}}
\newcommand{\een}{\end{enumerate}}

\newcommand{\bpm}{\begin{pmatrix}}
\newcommand{\epm}{\end{pmatrix}}

\title{The Seiberg-Witten K\"ahler Potential as a\\
Two-Sphere Partition Function}

\author{Daniel S. Park$^{1}$ and Jaewon Song$^2$\\
\\
$^1$Simons Center for Geometry and Physics\\
Stony Brook University\\
Stony Brook, NY 11794-3636, USA\\
\\
$^2$Department of Physics\\
University of California, San Diego\\
La Jolla, CA 92093, USA
\\
\\
{\tt dpark} {\rm at} {\tt scgp.stonybrook.edu},
{\tt jsong} {\rm at} {\tt physics.ucsd.edu}
}

\preprint{UCSD-PTH-12-15}

\abstract{Recently it has been shown that
the two-sphere partition function of a
gauged linear sigma model of a Calabi-Yau manifold
yields the exact quantum K\"ahler potential
of the K\"ahler moduli space of that manifold.
Since four-dimensional $\NN=2$ gauge theories
can be engineered by non-compact
Calabi-Yau threefolds, this implies that it is
possible to obtain exact gauge theory K\"ahler
potentials from two-sphere partition functions.
In this paper, we demonstrate that the
Seiberg-Witten K\"ahler potential can indeed be
obtained as a two-sphere partition function.
To be precise, we extract the quantum K\"ahler metric of
4D $\NN=2$ $SU(2)$ Super-Yang-Mills theory by taking the field
theory limit of the K\"ahler parameters of
the $\OO(-2,-2)$ bundle over $\field{P}^1 \times \field{P}^1$.
We expect this method of computing the K\"ahler potential
to generalize to other four-dimensional $\NN=2$ gauge
theories that can be geometrically engineered by toric Calabi-Yau
threefolds.}

\begin{document}

\section{Introduction and Summary} \label{s:intro}

Exact $S^2$ partition functions of two-dimensional
gauge theories with $(2,2)$ supersymmetry have recently
been computed using localization techniques
\cite{Benini:2012ui,Doroud:2012xw}.
By observing the properties of these partition functions
for gauged linear sigma models (GLSM's)
\cite{Witten:1993yc,Hori:2006dk,Donagi:2007hi,
Hori:2011pd,Jockers:2012zr} 
of Calabi-Yau threefolds, it was conjectured
\cite{Jockers:2012dk} ---
and later proven \cite{Gomis:2012wy} ---
that the $S^2$ partition computes
the exact quantum K\"ahler potential of the K\"ahler
moduli space of the manifold.
More precisely, the $S^2$ partition function
computes the inner-product
$\vev{0|\bar{0}}$, where $|0\rangle$ ($|\bar{0}\rangle$)
are the topological (anti-topological) ground states
of the A-twisted GLSM corresponding to the unit operator,
respectively \cite{Cecotti:1991me}.
In the case that the GLSM flows to a non-linear sigma
model (NLSM) of a Calabi-Yau manifold, this implies that
\be
Z_{S^2} = e^{-K} \,,
\label{conj}
\ee
where $Z_{S^2}$ is the $S^2$ partition
function of the GLSM and $K$ is the quantum
K\"ahler potential of the K\"ahler moduli space of
the Calabi-Yau manifold.

It follows that it should be possible to obtain the
K\"ahler potential of $\NN=2$
gauge theories in four dimensions by computing
$S^2$ partition functions, since these theories can be engineered
using non-compact Calabi-Yau threefolds
\cite{Kachru:1995fv,Katz:1996fh, Katz:1997eq}.
In particular, the gauge theory K\"ahler potential can be
obtained from the string partition function
on the corresponding Calabi-Yau threefold by
taking a limit of the K\"ahler parameters
of the threefold in which gravity decouples.
In this paper, we check that this is indeed
the case for 4D $SU(2)$ Super-Yang-Mills theory
(SYM) that is engineered by the non-compact
Calabi-Yau threefold
$\OO(-2,-2) \ra\field{P}^1 \times \field{P}^1$
\cite{Katz:1996fh}.
That is, we find that $K_{SW}$ ---
the Seiberg-Witten K\"ahler potential
of $\NN=2$ $SU(2)$ SYM \cite{Seiberg:1994rs} ---
can be obtained from $Z_{S^2}$ ---
the $S^2$ partition function of the GLSM
of $\OO(-2,-2) \ra\field{P}^1 \times \field{P}^1$ ---
when we take the K\"ahler parameters of the
manifold to the ``field theory limit"
\cite{Kachru:1995fv,Katz:1996fh, Katz:1997eq}.

Let us be precise.
The $\OO(-2,-2)$ bundle over
$\field{P}^1 \times \field{P}^1$
--- which we denote by $\MM$
through the rest of the paper ---
can be thought of as a blow-up of
an $A_1$ singularity trivially fibered
over a $\field{P}^1$.
In this picture, one of the $\field{P}^1$'s
of $\MM$ is the base and the other
is the fiber obtained
by resolving the $A_1$ singularity.
The manifold has two
quantum K\"ahler parameters
\be
t_b \equiv 2 \pi \xi_b - i \theta_b,\quad
t_f \equiv 2 \pi \xi_f - i \theta_f,
\ee
that control the size of the base/fiber
$\field{P}^1$, respectively.
These parameters, from the point of
view of the GLSM, are Fayet-Iliopoulos (FI)
parameters of the worldsheet
$U(1)$ gauge symmetries.
The $S^2$ partition function of
the GLSM are functions of these parameters.

There is, however, a subtlety in computing
the partition function of a GLSM that
is expected to flow to a non-linear sigma model
of a non-compact manifold in the infra-red limit.
Since there is a non-compact direction in the
geometry, the partition function is expected to diverge.
A way of controlling this divergence is to turn on a
small R-charge along the non-compact direction.
In this paper, we compute the $S^2$ partition function
of the GLSM of $\MM$ deformed by the R-charge
$\q$, {\it i.e.,}
\be
Z_{S^2} (t_b,t_f;\q) \,.
\ee

Meanwhile, the Seiberg-Witten K\"ahler potential
\cite{Seiberg:1994rs} of $\NN=2$ $SU(2)$ SYM is
a function of the gauge invariant modulus
\be
u = \vev{\Tr \, \phi^2}
\ee
--- where $\phi$ is the adjoint scalar field
in the $SU(2)$ vector multiplet ---
and the renormalization scale $\Lambda$,
{\it i.e.,}
\be
K_{SW} (u,\Lambda) \,.
\ee

The main result of this paper is that
\be
\lim_{\q \ra 0^+} \left[
 {1 \ov \q^2}
\lim_{\epsilon \ra 0}
\left( { -{\ln \epsilon \ov \epsilon^{2-4\q}}\p \bar{\p} \ln Z_{S^2} (t_b,t_f;\q) } \right) \right]
= - \pi \p \bar{\p} K_{SW}(u,\Lambda)
\label{mainresult}
\ee
if we identify
\begin{align}
\begin{split}
q_b &\equiv e^{-t_b} = \epsilon^4 \Lambda^4 \\
q_f &\equiv e^{-t_f} = {1 \ov 4} -\epsilon^2 u \,.
\end{split}
\label{ftlimit}
\end{align}
The partial derivatives in \eq{mainresult} are
taken with respect to $u$.
Equation \eq{ftlimit} defines the field theory limit
of the Calabi-Yau threefold. $\NN=2$ $SU(2)$
Super-Yang-Mills theory can be engineered
from Calabi-Yau manifold $\MM$ as its
K\"ahler parameters approach the point
\begin{align}
\begin{split}
q_b =0,\quad q_f ={1\ov 4}
\end{split}
\end{align}
in the moduli space with scaling \eq{ftlimit}
as $\epsilon \ra 0$.

Although we have restricted our attention to
$SU(2)$ Super-Yang-Mills theory in this paper,
we expect our strategy of computing the
quantum K\"ahler potential to be generalizable
to other $\NN=2$ gauge theories that can be
geometrically engineered from
toric Calabi-Yau threefolds \cite{Katz:1996fh, Katz:1997eq}.
If our expectations are correct,
the $S^2$ partition would provide 
yet another way \cite{Nekrasov:2002qd, Iqbal:2003ix, Iqbal:2003zz,
Hollowood:2003cv, Eguchi:2003sj}
of computing the K\"ahler potential for
such gauge theories.
We elaborate on issues related to generalizing
our computation to other theories at the
end of this paper.

This paper is organized as follows.
In section \ref{s:S2}
we summarize the calculation of
$Z_{S^2}$ and verify equation \eq{mainresult}.
In the process, we propose the R-charge
assignments of the fields of the GLSM
one needs to use to obtain the K\"ahler
potential of the non-compact Calabi-Yau
manifold $\MM$.
We use the correspondence
between A-twisted states of sigma models of
non-compact toric manifolds and compact
toric hypersurfaces presented in
\cite{Morrison:1994fr,Hori:2000kt}
to support our prescription.
We also discuss the field theory limit
\eq{ftlimit} in more detail.
In section \ref{s:gen} we lay out how
$S^2$ partition functions can be utilized to compute
the K\"ahler potential of other $\NN=2$ gauge theories
that can be engineered by toric Calabi-Yau threefolds
and describe issues that must be addressed along the way.
A detailed account of the calculation of $Z_{S^2}$
can be found in the appendix.

\section{The $S^2$ Partition Function and $SU(2)$
Super-Yang-Mills Theory} \label{s:S2}

In this section, we compute the $S^2$ partition function
for the gauged linear sigma model of $\MM$ in the field theory
limit, and extract the Seiberg-Witten K\"ahler
metric. In section \ref{ss:setup} we describe the
gauged linear sigma model of manifold $\MM$
and write down its $S^2$ partition function according to
\cite{Benini:2012ui,Doroud:2012xw}.
In particular, we argue that we must assign all chiral
fields of the GLSM to have R-charge zero in order for
it to flow to the sigma model of the non-compact Calabi-Yau
manifold $\MM$ in the infra-red limit.
In section \ref{ss:summ} we evaluate the $S^2$
partition function in the field theory limit \eq{ftlimit}
and verify that it is related to the Seiberg-Witten K\"ahler
metric \cite{Seiberg:1994rs} by \eq{mainresult}.

\subsection{The Setup} \label{ss:setup}

In this section, we write down the gauged linear sigma
model of $\MM$ and describe its field theory limit.
We then present the expression for the $S^2$
partition function and justify the choices of R-charges
we use for the chiral fields of the GLSM.
Most of the contents of this section are thoroughly
explained, among other places,
in \cite{Benini:2012ui,Doroud:2012xw,Witten:1993yc,
Jockers:2012dk,Gomis:2012wy,Kachru:1995fv,Katz:1996fh}
--- we have merely stated them in a way
that is convenient for our purposes.

The two-dimensional gauged linear sigma model
of $\MM$ --- the $\OO(-2,-2)$
bundle over $\field{P}^1 \times \field{P}^1$ --- is
given by a $\mathcal{N}=(2, 2)$ supersymmetric gauge theory
with gauge group $U(1) \times U(1)$ and five chiral
multiplets.\footnote{Gauged linear sigma models for
various manifolds have been constructed in
\cite{Witten:1993yc,Hori:2006dk,Donagi:2007hi,
Hori:2011pd,Jockers:2012zr}. For a pedagogical
review of GLSM's, see \cite{Denef:2008wq}.}
The charges of the chiral multiplets under the gauge group is
summarized in table \ref{t:chiral}.
The two $U(1)$'s each correspond to the $\field{C}^*$ action
that acts on the projective coordinates of each $\field{P}^1$.

It is useful to think of $\MM$ as a resolution of
an $A_1$ singularity fibered over
a $\field{P}^1$. Let us denote one of the $\field{P}^1$'s as
the base and the other as the fiber.
The real part of the two FI parameters
for each $U(1)$ gauge symmetry
\be
t_b \equiv 2 \pi \xi_b - i \theta_b,\quad
t_f \equiv 2 \pi \xi_f - i \theta_f,
\ee
control the size of the base and fiber $\field{P}^1$ respectively.
These FI parameters are the two quantum K\"ahler parameters
of the Calabi-Yau threefold.

$\NN=2$ $SU(2)$ Super-Yang-Mills theory is geometrically
engineered by ``compactifying" type IIA string theory on this
manifold and then by taking the K\"ahler parameters to
a certain limit \cite{Kachru:1995fv,Katz:1996fh,Katz:1997eq}.
In order to decouple gravity, one must take the size of the
base --- controlled by $t_b$ --- to be infinite,
but at the same time scale the size of the
fiber --- controlled by $t_f$ --- appropriately
so that the gauge coupling is finite.
In this scaling limit, $t_b$ sets the renormalization
scale of the gauge theory while $t_f$ controls the mass
of the $W$ bosons, as the $W$ bosons come from
quantizing D2-branes wrapping the fibral $\field{P}^1$
\cite{Strominger:1995ac,Witten:1996qb}.
We can parametrize the mass of the $W$ bosons by
the gauge invariant Coulomb branch parameter
\be
u = \vev{\Tr \, \phi^2}
\ee
where $\phi$ is the adjoint scalar field in the 
$\NN=2$ $SU(2)$ vector multiplet.

The field theory limit can be obtained by setting
\begin{align}
q_b &\equiv e^{-t_b} = \epsilon^4 \Lambda^4
\end{align}
as in \cite{Katz:1996fh}, and by taking $\epsilon$ to
be small. It turns out that one must take
\begin{align}
q_f &\equiv e^{-t_f} = {1 \ov 4} -\epsilon^2 u
\end{align}
accordingly.
Note that this is the same scaling limit discussed in \cite{Lawrence:1997jr}. 
One might have expected $u$ to parametrize the
deviation of the geometry from the point $t_f=0$
--- when the fibral $\field{P}^1$
shrinks to zero size, where the full $SU(2)$ symmetry
of the classical theory is presumably recovered.
We, however, find that the
``classical singular point" one must expand around in
our case is located at $q_f =1/4$.\footnote{By
``classical singular point" we refer to a singular point in
the classical moduli space of the four-dimensional gauge
theory, which in our case is $SU(2)$ SYM. We note that
this point is not singular with respect to the quantum metric
on the moduli space \cite{Seiberg:1994rs}. Instead, the
quantum moduli space develops new singularities where
the monopoles or dyons become massless. }
As we show in the next section,
the quantum K\"ahler potential computed by the
$S^2$ partition function is written as an expansion in
\be
{q_b \ov (1-4q_f)^2} \,.
\ee
Since we know that the instanton contributions to
the K\"ahler potential of the $SU(2)$ theory come in
powers of $\Lambda^4/u^2$ \cite{Seiberg:1994rs,Seiberg:1988ur},
the assignment \eq{ftlimit} is a natural starting point. 

\begin{table}[t!]
\center
  \begin{tabular}{ | c | c | c | c | c | c | }
  \hline
  &$B_1$ & $B_2$ & $F_1$ & $F_2$ & $P$ \\ \hline
  $U(1)_b$ & 1 & 1  & 0 & 0 & $-2$ \\ \hline
  $U(1)_f$  & 0  & 0 & 1 & 1 & $-2$ \\ \hline
    \end{tabular}
  \caption{Charges of the five chiral multiplets of the GLSM of the
  $\OO(-2,-2)$ bundle over $\field{P}^1 \times \field{P}^1$.
  The chiral multiplets $B_i$ can be thought of as projective coordinates
  of the base $\field{P}^1$, while $F_i$ can be thought of those of the fiber.
  $P$ is the $\OO(-2,-2)$ bundle coordinate.}
\label{t:chiral}
\end{table}

Now we are almost in a position to compute the field theory limit
of the $S^2$ partition function of the GLSM for $\MM$.
The $S^2$ partition function of a two-dimensional
$\NN=(2,2)$ abelian gauge theory with gauge group
$U(1)^N$ and $A$ chiral multiplets can be computed
as follows.\footnote{We do not elaborate on the beautiful
physics behind the computation of $Z_{S^2}$ here
as it has already been explained eloquently in the
original papers \cite{Benini:2012ui,Doroud:2012xw}.}
In the absence of a twisted superpotential, the partition
function obtained via localization is given by
\be
\sum_{\{m_n \} \in \field{Z}^N} e^{-i\sum_n m_n \theta_n}
 \prod_n \left(\int_{-\infty}^\infty {d\sigma_n \ov 2\pi} \right) 
e^{-4\pi i \sum_n \xi_n \sigma_n}
\prod_a Z_a (\{ \sigma_n \}, \{ m_n \};\q_a,\{ q_{an} \})
\label{Z}
\ee
where the $n$ index runs over $1, \cdots, N$
and the $a$ index runs over $1, \cdots, A$.
We have used $\{ x_n \}$ as a shorthand
notation for the $N$-tuple $(x_1,\cdots,x_N)$.
The parameters $2 \pi \xi_n$ and $-\theta_n$ are the real
and imaginary part of the FI parameter
\be
t_n = 2 \pi \xi_n - i\theta_n
\ee
of $U(1)_n$.
$Z_a$ is a function of the parameters $\{ \sigma_n \}, \{ m_n \}$
as well as the charges $\q_a$ and $q_{an}$ of the $a$'th
chiral multiplet $\Phi_a$.
We have used $q_{an}$ to denote the charge of
$\Phi_a$ under the $n$'th gauge group $U(1)_n$,
and $\q_a$ to denote its R-charge. $Z_a$ is then given by
\be
Z_a = {\Ga({\q_a \ov 2}-i \sum_n q_{an} \sigma_n -  \ha \sum_n q_{an} m_n  )
\ov \Ga(1-{\q_a \ov 2}+i \sum_n q_{an} \sigma_n -  \ha \sum_n q_{an} m_n  )} \,.
\ee

What remains to be understood is what the R-charges of the
chiral fields of the GLSM should be in order for it to flow to
$\MM$ in the infra-red limit of the theory. A naive prescription
would be to assign all R-charges to be zero, as this would prevent
any kind of superpotential from being written down.
We claim that this naive prescription is correct.

Our claim follows directly from \cite{Hori:2000kt}
since $Z_{S^2}$ computes the inner-product between
$|0\rangle$ and $|\bar{0}\rangle$ ---
the topological and anti-topological ground states
of the A-twisted GLSM corresponding to the unit operator
\cite{Jockers:2012dk,Gomis:2012wy}.
To elaborate, our prescription follows from the relation
between A-twisted ground states of compact and
non-compact Calabi-Yau manifolds discussed in
\cite{Hori:2000kt}. Let us illustrate this relation
in our case.

The $U(1)^2$ gauge theory with the matter
content given in table \ref{t:chiral} flows
to a NLSM on a compact Calabi-Yau
hypersurface of
$\field{P}^1 \times \field{P}^1$ --- which
we denote by $\MM_c$ --- in the presence
of a gauge invariant superpotential
\be
P G(B_i, F_i) \,,
\ee
in the ``geometric phase", {\it i.e.,}
when $t_b$ and $t_f$ are large.
Here $G$ is a generic polynomial of homogenous
degree two in each of the $B$ and $F$ fields.
In order for the superpotential of this theory to be of this form,
the $R$ charges of the chiral fields must be given as in
table \ref{t:chiralR}.
The $S^2$ partition function of this compact theory
is computed according to the formula \eq{Z}.
As elaborated in \cite{Jockers:2012dk}, if one assumes that
the $S^2$ partition is the exponential of the K\"ahler
metric of the moduli space as in \eq{conj}
\be
Z_{S^2} = e^{- K} \,,
\nonumber
\ee
one can observe that varying the charge assignments
$b$ and $f$ only act as K\"ahler transformations
\be
K \ra K + f + \bar{f}
\ee
on the K\"ahler potential ---
it does not affect the computation of the quantum
K\"ahler metric. This should be the case, as the R-charge of
all gauge invariant operators stay invariant under
different assignments of $b$ and $f$ in table \ref{t:chiralR}.
Therefore one may take these charges to have a very small, but positive
value. The $S^2$ partition function is then given by
\begin{align}
\begin{split}
Z^c_{S^2} &= \sum_{m_f, m_b \in \field{Z}} e^{-im_f \theta_f -im_b \theta_b}
\int {d\sig_b d\sig_f \ov (2\pi)^2}
e^{-4\pi i \xi_f \sig_f -4\pi i \xi_b \sig_b} \\
&\times {\Ga (-i\sig_f - {m_f \ov 2})^2 \ov \Ga (1+i\sig_f - {m_f \ov 2})^2}
 {\Ga (-i\sig_b - {m_b \ov 2})^2 \ov \Ga (1+i\sig_b - {m_b \ov 2})^2}
 {\Ga (1+2i\sig_b+2i\sig_f +m_b+m_f) \ov \Ga (-2i\sig_b-2i\sig_f +m_b+m_f)} \,,
\end{split}
\label{ZS2_c}
\end{align}
where the contour of integration for $\sigma_b$ and
$\sigma_f$ are along the real axis with a small
``jump" over the origin. The jump is made to push the
poles located at the origin below the contour of integration,
as we are assigning small positive R-charges to the fields
$B_i$ and $F_i$.

\begin{table}[t!]
\center
  \begin{tabular}{ | c | c | c | c | c | c | }
  \hline
  &$B_1$ & $B_2$ & $F_1$ & $F_2$ & $P$ \\ \hline
  $\MM_c$ & $b$ & $b$  & $f$ & $f$ & $2-2b-2f$ \\ \hline
    \end{tabular}
  \caption{R-charge assignments of the chiral fields used
  for computing the quantum K\"ahler potential of the
  manifold $\MM_c$.}
\label{t:chiralR}
\end{table}

The A-twisted state $|0 \rangle_c$ of the compact
theory on $\MM_c$ and the A-twisted state $|0 \rangle$
of the non-compact theory on $\MM$ are related by
\cite{Hori:2000kt}
\be
|0 \rangle_c = \delta |0 \rangle
\ee
where the operator $\delta$ is defined to be
\be
\delta = 2 \Sigma_b + 2 \Sigma_f \,.
\ee
$\Sigma_b$ and $\Sigma_f$ are the field strengths
of the gauge groups $U(1)_b$ and $U(1)_f$ respectively.
Therefore the inner-product between the topological and
anti-topological A-twisted states of the compact theory and the
non-compact theory are related by
\be
{}_c\vev{0 |\bar 0}_c  =
\langle 0| \delta \bar{\delta} |\bar{0} \rangle \,.
\ee
Now the right-hand-side of this equation can be
obtained from $\vev{0 |\bar 0}$ by taking two
derivatives with respect to the FI parameter
$T\equiv (2t_b +2t_f)$, {\it i.e.,}
\be
Z_{S^2}^c=\langle 0| \delta \bar{\delta} |\bar{0} \rangle \propto
{\p^2 \ov \p T \p \bar{T}} \langle 0|\bar{0} \rangle \,.
\label{rel}
\ee
It is easy to verify that the expression
\begin{align}
\begin{split}
Z_{S^2} &= \sum_{m_f, m_b \in \field{Z}} e^{-im_f \theta_f -im_b \theta_b}
\int {d\sig_b d\sig_f \ov (2\pi)^2}
e^{-4\pi i \xi_f \sig_f -4\pi i \xi_b \sig_b} \\
&\times {\Ga (-i\sig_f - {m_f \ov 2})^2 \ov \Ga (1+i\sig_f - {m_f \ov 2})^2}
 {\Ga (-i\sig_b - {m_b \ov 2})^2 \ov \Ga (1+i\sig_b - {m_b \ov 2})^2}
 {\Ga (2i\sig_b+2i\sig_f +m_b+m_f) \ov \Ga (1-2i\sig_b-2i\sig_f +m_b+m_f)} \,,
\end{split}
\label{ZS2_1}
\end{align}
with the contour of integration for $\sigma_b$ and
$\sigma_f$ along the real axis with a small
jump over the origin satisfies
\be
Z^c_{S^2} \propto {\p^2 \ov \p T \p \bar{T}} Z_{S^2} \,,
\label{rel2}
\ee
upon comparison with \eq{ZS2_c}.
The equation \eq{ZS2_1} comes from setting the
R-charges of all the chiral fields to be zero,
and hence our claim.
We note that proportionality constants in the equations
\eq{rel} and \eq{rel2} are irrelevant when computing the
quantum K\"ahler metric of the moduli space and therefore
can be ignored.

There is one more issue that must be addressed when
computing the $S^2$ partition function for the GLSM of $\MM$.
There is an ambiguity in computing the $S^2$ partition function
\eq{ZS2_1} that must be resolved.
Unlike in the compact case, the third factor
of the integrand becomes singular at the origin of the complex
$\sigma_b$ and $\sigma_f$ plane. One must be careful how
to ``split the poles" of the integrand at the origin, as it affects
the value of the integral. The correct way to deal with this is to
assign a small positive R-charge to all the chiral fields and take
the contour of integration to be along the real axis \cite{Benini:2012ui}.
An efficient way to implement this prescription is to assign a
small positive R-charge $2\q$ to the field $P$, and take the
contour of integration for $\sigma_b$ and
$\sigma_f$ to be along the real axis with a small
``jump" over the origin.
In this case, the R-charges of all gauge invariant operators
become positive multiples of $2 \q$ and hence positive
--- this is a necessary condition for the theory to be unitary
\cite{Benini:2012ui,Doroud:2012xw}.
The $S^2$ partition function for $\MM$ can be recovered in
the $\q \ra 0^+$ limit.

We can motivate introducting the R-charge $\q$ in another way.
$\MM$ is a non-compact manifold and hence has a divergent
direction. We see in the next section that indeed
\eq{ZS2_1} is divergent in the geometric phase.
In order to compute the quantum K\"ahler potential of this
theory, we must regulate this divergence.
This task can be achieved by giving the $P$ field a
small R-charge $2\q$
and observing the behavior of $Z_{S^2}$ as we take
$\q$ to be small.

We therefore compute the $S^2$ partition function
\begin{align}
\begin{split}
Z_{S^2} &= \sum_{m_f, m_b} e^{-im_f \theta_f -im_b \theta_b}
\int_{-\infty+i0^+}^{\infty+i0^+} {d\sig_b d\sig_f \ov (2\pi)^2}
e^{-4\pi i \xi_f \sig_f -4\pi i \xi_b \sig_b} \\
&\times {\Ga (-i\sig_f - {m_f \ov 2})^2 \ov \Ga (1+i\sig_f - {m_f \ov 2})^2}
 {\Ga (-i\sig_b - {m_b \ov 2})^2 \ov \Ga (1+i\sig_b - {m_b \ov 2})^2}
 {\Ga (\q+2i\sig_b+2i\sig_f +m_b+m_f) \ov \Ga (1-\q-2i\sig_b-2i\sig_f +m_b+m_f)}
\end{split}
\label{ZS2}
\end{align}
for small positive $\q$ in the field theory limit
\eq{ftlimit}
\begin{align}
\begin{split}
q_b &\equiv e^{-t_b} = \epsilon^4 \Lambda^4 \nonumber \\
q_f &\equiv e^{-t_f} = {1 \ov 4} -\epsilon^2 u \,. \nonumber
\end{split}
\end{align}
The contour of integration is taken to be slightly
above the real axes of the complex $\sigma_b$
and $\sigma_f$ planes.
As we see in the next section,
the quantum K\"ahler potential of $SU(2)$
SYM can be extracted from this partition function
by examining its the leading order behavior of 
in $\epsilon$ and $\q$.

\subsection{Summary of Calculation} \label{ss:summ}

In this section, we summarize the calculation of the
$S^2$ partition function \eq{ZS2} of the gauged
linear sigma model for $\MM$.
We have presented a detailed account
of the calculation in the appendix.
We comment on peculiar aspects of $Z_{S^2}$
when we take $\q <0$ at the end of this section.

We evaluate \eq{ZS2} in the field theory limit
\eq{ftlimit} in two steps following \cite{Katz:1996fh}:
\begin{enumerate}
\item We first evaluate \eq{ZS2} in the large volume ---
or geometric --- limit, where we take the K\"ahler
parameters of the manifold to be large.
The partition function in this limit is written as an
expansion of $q_f$ and $q_b$ defined in \eq{ftlimit},
as these parameters are small in the large volume
limit.
\item We complete the sum over $q_f$ and take the
field theory limit to obtain the gauge theory K\"ahler
potential.
\end{enumerate}
Let us explain the second step in more detail.
To arrive at the field theory limit, we must take the
base of the manifold to be large --- {\it i.e.,}
take $q_b$ to be small --- but at the same
time shrink the fiber near a classical singular point
in the moduli space.
The limit of taking the base
to be large, more precisely taking \cite{Katz:1996fh}
\be
q_b =\epsilon^4 \Lambda^4
\ee
for small $\epsilon$ is compatible with the
large volume limit.
The field theory limit for the size of the fiber, however,
is more subtle. By summing the series with respect
to $q_f$ in the large volume limit, $Z_{S^2}$ can be
written in the form
\begin{align} \label{s2su2result}
\begin{split}
Z_{S^2} = {d^2 \ov d\alpha d\beta}
\Biggl[ &
\left( {\pi \alpha \ov \sin \pi \alpha}\right)^2
\left( {\pi \beta \ov \sin \pi \beta}\right)^2
(q_f \bar{q}_f)^{-\alpha}
(q_b \bar{q}_b)^{-\beta}
\left( {\sin \pi(\q-2\alpha-2\beta) \ov \pi }\right) \\
& \left( \sum_{n_f \geq 0} {f_{n_f-\beta+{\q \ov 2},\alpha}(\bar{q}_f) \ov \Ga(1+n_f -\beta)^2}
\bar{q}_b^{n_f} \right)
\left( \sum_{p_f \geq 0} {f_{p_f-\beta+{\q \ov 2},\alpha}({q}_f) \ov \Ga(1+p_f -\beta)^2}
{q}_b^{p_f} \right)
\Biggr]_{\alpha=\beta=0} \,,
\end{split}
\end{align}
where $f_{p_f-\beta+\q/2,\alpha}$ is a hypergeometric
function defined in \eq{fza}. Note that \eqref{s2su2result}
is symmetric under the exchange of $q_b$ and $q_f$ as expected. 
The function $f_{p_f-\beta+\q/2,\alpha}$ has a singularity at
$q_f =1/4$, {\it i.e.,}
\be
f_{p_f-\beta+\q/2,\alpha} \propto (1-4q_f)^{1/2-2p_f-\q+2\beta}
\ee
as $q_f \ra1/4$. It is clear that when $q_b$ is small and
$q_f$ is near $1/4$, $Z_{S^2}$ has a series expansion in
\be
{q_b \ov (1 -4q_f)^2}
\ee
at leading order in these small parameters.
Therefore we see that $q_f = 1/4$ is the classical
singular point of the gauge theory --- the individual terms
of the instanton expansion are singular around this point.
It follows that the Coulomb branch parameter $u$
of the gauge theory should parametrize the deviation
of a point in the moduli space from this classical
singular point. The correct gauge theory limit is
thus obtained by
\be
q_f ={1 \ov 4} - \epsilon^2 u \,,
\ee
where the $S^2$ partition function has
an expansion in $\Lambda^4/u^2$, as desired
\cite{Seiberg:1994rs}. 

It is interesting that the two-sphere partition
function is more naturally written in terms of
the ``mirror coordinate" $u$ which parametrizes
the complex structure of the Seiberg-Witten curve
--- the mirror geometry of the manifold $\MM$
 \cite{Katz:1996fh}.
An analogous observation can be made about
GLSM's of compact Calabi-Yau threefolds as well
\cite{Jockers:2012dk}.
In this reference, the authors compute Gromov-Witten
invariants of compact Calabi-Yau threefolds
from the $S^2$ partition function.
In order to do so, a coordinate transform is implemented.
The Gromov-Witten invariant of a compact Calabi-Yau
manifold $M$ can be extracted by
using a different set of K\"ahler coordinates
$q_l' =e^{-t'_l}$ from $q_l =e^{-t_l}$, which is
built into the gauged linear sigma model of $M$
--- recall that $t_l$ are the FI parameters
of the two-dimensional gauge theory.
It turns out that the relation between
$t_l$ and $t_l'$ is given by the mirror map
--- the K\"ahler coordinates built into
the GLSM have a natural interpretation in terms
of the classical geometry of the mirror manifold
of $M$, rather than that of the manifold $M$ itself.
In our example this ``mirror map" gives the relation
between the ``IR" coordinate $u$ and the
``UV" coordinate $a$, where $a$ is the
vacuum expectation value of the
adjoint scalar $\langle \phi \rangle = \textrm{diag}(a, -a)$
in the $SU(2)$ vector multiplet.

Evaluating the integral for $Z_{S^2}$ in
the field theory limit,
the small $\epsilon$ expansion of the
full partition function can be written as
\be
Z_{S^2} = Z_L \ln \epsilon + Z_0 +
\epsilon^{2-4\q} Z_{2-4\q} + \OO(\epsilon^2) \,.
\label{ZS2exp}
\ee
Since $\q$ is positive, we can conveniently isolate
$Z_{2-4\q}$ apart from the order $\epsilon^2$ terms.
In the $\q \ra 0^+$ limit --- where the GLSM becomes
non-compact --- the terms $Z_L$ \eq{ZL} and
$Z_0$ \eq{Z0} diverge as
\be
Z_L \sim -{16 \ov \q^2},\quad
Z_0 \sim {8 \ov \q^3} \,,
\ee
while $Z_{2-4\q}$ approaches the Seiberg-Witten
K\"ahler potential
\be
Z_{2-4\q} \sim -16 \pi K_{SW} \,.
\ee
We can thus observe the divergent behavior of
$Z_{S^2}$ in the non-compact limit.

The quantum K\"ahler potential of the theory ---
according to \cite{Jockers:2012dk,Gomis:2012wy}
--- is given by
\be
-\ln Z_{S^2} =
-\ln (Z_L \ln \epsilon+Z_0)
- {\epsilon^{2-4\q} \ov \ln \epsilon}
{Z_{2 -4\q} \ov Z_L} + 
\text{(sub-leading in $\epsilon$)} \,.
\ee
This near-field theory --- or small $\epsilon$ ---
behavior of the K\"ahler potential is consistent with
that observed in mirror symmetry calculations
in the field theory limit
\cite{Billo:1997vt,Eguchi:2007iw}.
The leading term drops out when computing
the quantum K\"ahler metric of the gauge theory moduli
space, as both $Z_L$ and $Z_0$ are
independent of $u$. Hence the $Z_{2-4\q}$ term
is the leading term in the $\epsilon \ra 0$ limit
for the K\"ahler metric:
\be
-\p_u \p_{\bar{u}} \ln Z_{S^2} =
- {\epsilon^{2-4\q} \ov \ln \epsilon}
{\p_u \p_{\bar{u}} Z_{2 -4\q} \ov Z_L} + 
\text{(sub-leading in $\epsilon$)} \,.
\ee
We must eventually take $\q$ to zero to recover the
partition function for the non-compact manifold $\MM$.
As shown in the appendix
\be
-\lim_{\epsilon \ra 0} 
\left( - {\ln \epsilon \ov {\epsilon^{2-4\q}} }\right)
 \p_u \p_{\bar{u}} \ln Z_{S^2}
= {\pi \q^2  }  \p_u \p_{\bar{u}} K_{SW} (u,\Lambda)
+ \text{(sub-leading in $\q$)} \,,
\ee
and hence our main result.

Let us end this section by commenting on the behavior
of the $S^2$ partition function when $\q$ is negative,
{\it i.e.,} when we split the pole of the integrand of \eq{ZS2_1}
in a different way. In this case, one must add to \eq{ZS2exp}
terms that come from poles that are ``pushed below"
the contour of integration by taking $\q$ to be negative.
$Z_{S^2}$ can be written as
\be
Z_{S^2} = Z_0' + \epsilon^{-4q} Z_{-4q}'
+\epsilon^{2} Z_{2}' + \epsilon^{2-4\q} Z_{2-4\q}' +
(\text{sub-leading in $\epsilon$}) \,.
\ee
It turns out that there is a rather miraculous cancellation of
singularities such that the behavior of the leading term
in this expansion is given by
\be
Z_0' \ra -14 \zeta(3)
\ee
as $\q \ra 0$. The subleading term
$Z_{-4q}$, however, is divergent in this limit.
We note that $Z_{2-4\q}'$ is modified from $Z_{2-4\q}$
by terms that can be written as a sum of
holomorphic and anti-holomorphic functions of $u$.
Not only is the $S^2$ partition function negative for
$\epsilon \ra 0$ in this case, but also the K\"ahler metric
becomes negative definite at leading order in $\epsilon$.
These pathologies should not come as
a surprise, as the gauge invariant operators of the theory
with negative $\q$ have negative R-charge ---
such theories are not unitary.

\section{Generalization to Other $\NN=2$ Gauge Theories} \label{s:gen}

Since the $S^2$ partition can compute the exact
quantum K\"ahler potential of the K\"ahler moduli
space for non-compact Calabi-Yau manifolds
that have a two-dimensional GLSM, it should be possible
to use it for computing the quantum K\"ahler potential for
more general gauge theories.
In particular, our computations should generalize
to $\NN=2$ gauge theories that can be engineered
by toric Calabi-Yau threefolds
\cite{Katz:1996fh,Katz:1997eq,Iqbal:2003ix,
Iqbal:2003zz,Hollowood:2003cv}.
More precisely, we expect that the quantum K\"ahler
potential of a gauge theory $T$ geometrically
engineered by a toric Calabi-Yau threefold
$M$ can be obtained through the following steps:
\begin{enumerate}
\item Construct the gauged linear sigma model of $M$ and
deform the theory by small R-charges $\q_i$ to
regulate the non-compact directions.
\item Identify the correct field theory limit parametrized by
small parameters $\epsilon_a$.
\item Compute the $S^2$ partition function $Z_{S^2}$
of the gauged linear sigma model.
\item Obtain the gauge theory K\"ahler potential --- defined up
to K\"ahler transformations --- by observing the leading order
behavior of $Z_{S^2}$ in the small parameters $\q_i$ and
$\epsilon_a$.
\end{enumerate}
These steps straightforwardly follow from
geometric engineering and the results of
\cite{Jockers:2012dk,Gomis:2012wy}.
There are, however, some details that need to
be worked out in order to actually follow through
with the computation.

As we have worked with the simplest $\NN=2$
theory in this paper, the field theory
limit of the K\"ahler parameters were identified
with relative ease. For theories
coming from more complicated manifolds, the
field theory limit of the manifold must be worked
out with greater care
\cite{Katz:1996fh,Katz:1997eq,Iqbal:2003ix, Iqbal:2003zz,
Hollowood:2003cv}.
Also, the task of identifying K\"ahler parameters
with physical observables is more involved.
Let us illustrate these issues
through the example of pure $SU(N)$ SYM.
The Calabi-Yau threefold $M$ that engineers
$\NN=2$ $SU(N)$ SYM is given by
an $A_{N-1}$ singularity fibered over $\field{P}^1$.
We have $N-1$ K\"ahler parameters $t_f^i$, $i=1, \cdots, N-1$
for the fiber and one parameter $t_b$ for the base
\cite{Katz:1996fh}.
The field theory limit of the base coordinate,
as in the case of $SU(2)$, can be easily shown to be
\be
 q_b = e^{-t_b} \sim (\e \Lambda)^{b_0} \,.
\ee
The exponent of this equation is given by the
coefficient of the beta function $b_0 = 2N$.
In the meanwhile, finding the correct field theory
limit for the fiber K\"ahler parameters
\be
q_f^i = e^{-t_f^i}
\ee
is more involved. In order to find the scaling limit
for the fiber coordinates, we must first identify
the point in the moduli space we must expand around.
Since the GLSM picks out the K\"ahler coordinates
we must use, identifying this point is not trivial.
Also, as demonstrated in the case of $SU(2)$,
the quantum K\"ahler coordinates given by the
gauged linear sigma model are
naturally written in terms of the ``IR"
parameters --- {\it i.e.,}
the gauge invariant operators ---
in the field theory limit.
The gauge invariant operators
parameterizing the Coulomb branch of $SU(N)$ SYM
are given by $u_n = \langle \Tr \phi^n \rangle$
with $n=2, \cdots, N$.
Finding the relation between the parameters
$u_n$ and the scaling limit of the $(N-1)$
K\"ahler parameters $q_f^i$ requires
some more work compared to the case of $SU(2)$.

There is also an issue with constructing the
GLSM for a given manifold $M$ ---
one must find the correct R-charges
to assign to the chiral fields of the two-dimensional
theory. For toric Calabi-Yau
manifolds with associated
compact Calabi-Yau hypersurfaces in the sense
of \cite{Hori:2000kt}, our argument for setting all
R-charges of the chiral fields to zero holds.
That is, if there exist chiral fields $P_\beta$,
$\beta=1,\cdots,l$ for the GLSM of $M$
such that for certain R-charge
assignments the most generic gauge-invariant
superpotential is given by the form
\be
\sum_\beta P_\beta G_\beta (X_i) \,,
\ee
we can repeat the argument of section
\ref{ss:setup} to show that all R-charges of the
chiral fields should be set to zero.
In this case, regulating the non-compactness is
also straightforward --- one can assign positive
R-charges $2\q_\beta$ for each $P_\beta$
and eventually take the $\q_\beta \ra 0^+$ limit.
For more general toric Calabi-Yau threefolds,
we do not know how to argue that such
a prescription works.
It would be interesting to understand which
R-charges to assign to the chiral fields
of the GLSM of a non-compact Calabi-Yau
threefold in general.

\section*{Acknowledgement}

We would like to thank Francesco Benini,
Nikolay Bobev, Koushik Balasubramanian, Peng Gao,
Jaume Gomis, Ken Intriligator, Vijay Kumar, Sungjay Lee
and David Morrison for useful discussions.
We would especially like to thank Francesco Benini
and Vijay Kumar for insightful comments on
the original draft of this paper.
We would also like to thank the Simons Center for
Geometry and Physics and the organizers of the
2012 Summer Simons Workshop in Mathematics
and Physics for their hospitality while this work
was being conceived.
DP thanks Wati Taylor for his support and
encouragement on this work.
DP is supported in part by DOE grant
DE-FG02-92ER-40697. The work of JS is supported
by DOE-FG03-97ER40546.
 
\appendix

\section{Evaluation of the Partition Function} \label{ap:eval}

In this appendix, we evaluate the integral $Z_{S^2}$
defined in \eq{ZS2} ---
\begin{align}
\begin{split}
Z_{S^2} &= \sum_{m_f, m_b} e^{-im_f \theta_f -im_b \theta_b}
\int {d\sig_b d\sig_f \ov (2\pi)^2}
e^{-4\pi i \xi_f \sig_f -4\pi i \xi_b \sig_b} \\
&\times {\Ga (-i\sig_f - {m_f \ov 2})^2 \ov \Ga (1+i\sig_f - {m_f \ov 2})^2}
 {\Ga (-i\sig_b - {m_b \ov 2})^2 \ov \Ga (1+i\sig_b - {m_b \ov 2})^2}
 {\Ga (\q+2i\sig_b+2i\sig_f +m_b+m_f) \ov \Ga (1-\q-2i\sig_b-2i\sig_f +m_b+m_f)}
\end{split}
\nonumber
\end{align}
--- and compute its leading order behavior
in the field theory limit \eq{ftlimit} as we take $\epsilon \ra 0$.
As previously stated, we define the contour
of integration on the $\sig_f$ and $\sig_b$ planes to be
slightly above the real axis to avoid the pole lying on the
real axes. Recall that $\q$ is a small positive number,
{\it i.e.,}
\be
0 < \q \ll 1 \,.
\ee

Let us reiterate our strategy of computing $Z_{S^2}$
in the field theory limit.
We first compute the integrand in the large volume limit,
{\it i.e.,} when $\xi_b$ and $\xi_f$ are large.
The field theory limit involves taking $\xi_b$ to be large.
The problem is that we must eventually take a
small $\epsilon$ expansion
around a finite value of $t_f =2\pi \xi_f -i\theta_f$.
We do so by first summing the full expansion for $Z_{S^2}$
valid in the large $t_f$ limit --- more precisely a $q_f = e^{-t_f}$
expansion --- to obtain an expression valid for a generic value
of $t_f$. Then we continue this expression around the point
$q_f = {1/4}$.

We note one useful fact before evaluating the integral expression
for $Z_{S^2}$. As we present shortly,
we evaluate this integral by deforming the contour
of integration to the lower-half of the complex $\sig_b$
and $\sig_f$ planes and picking up poles of the integrand.
We note that one may obtain the final integral by only considering
poles with respect to $\sig_f$ and $\sig_b$ coming from the factors
\be
{\Ga (-i\sig_f - {m_f \ov 2})^2 \ov \Ga (1+i\sig_f - {m_f \ov 2})^2}
\quad \text{and} \quad
 {\Ga (-i\sig_b - {m_b \ov 2})^2 \ov \Ga (1+i\sig_b - {m_b \ov 2})^2}
\label{polecont}
\ee
when $\q$ is a small positive number.
This statement is due to the fact that the integral \eq{ZS2} picks
up codimension-two poles of the integrand.
All the codimension-two poles of the integrand lying in
the relevant region --- the product of the lower-half of
the two complex planes --- coincide with the contribution
from the two terms \eq{polecont}.

Let us begin the evaluation of $Z_{S^2}$
by completing the $\sig_f$ integral and compute
\begin{align}
\begin{split}
 \sum_{m_f} e^{-im_f \theta_f }
\int {d\sig_f \ov 2\pi}
e^{-4\pi i \xi_f \sig_f }
 {\Ga (-i\sig_f - {m_f \ov 2})^2 \ov \Ga (1+i\sig_f - {m_f \ov 2})^2}
 {\Ga (\q+2i\sig_b+2i\sig_f +m_b+m_f)
 \ov \Ga (1-\q-2i\sig_b-2i\sig_f +m_b+m_f)} \,.
\end{split}
\label{fint1}
\end{align}
Taking the large $\xi_f$ limit, we may deform the contour
of integration downwards on the complex $\sig_f$ plane
due to the exponential factor $e^{-4\pi i \xi_f \sig_f }$ in the integrand.
The integral becomes a sum of the residues of the poles
of the integrand in the lower-half of the complex $\sig_f$ plane.
As mentioned earlier, we only need to be concerned with
the poles due to the gamma function $\Ga (-i\sig_f - {m_f \ov 2})$
in the denominator. These are at the loci
\be
-i\sig_f-{m_f \ov 2} = -n_f, \qquad n_f \in \ZZ_{\geq 0}
\quad \text{and} \quad n_f \geq m_f \,.
\label{floci}
\ee
We note that $n_f \geq m_f$ since when
$n_f < m_f$ the pole of
$\Ga (1+i\sig_f - {m_f \ov 2})$ cancels this pole.

If the lowest degree of the
Laurent expansion of the function $f(z)$ around $z=z_0$
is $-n$, the residue of $f(z)$ at $z=z_0$ can be found by
\be
\Res{z=z_0} f(z) ={1 \ov \Ga(n)} {d^{n-1} \ov d\alpha^{n-1}}
(\alpha^{n} f(z_0+\alpha)) |_{\alpha=0} \,.
\label{fz}
\ee
Since the lowest degree of the Laurent expansion of
the integrand at the loci \eq{floci} is $-2$,
we can sum the residues of the integrand of
\eq{fint1} to obtain\footnote{We have actually used
a version of the equation \eq{fz} by replacing $\alpha \ra i\alpha$.}
\begin{align} 
\begin{split}
 \sum_{n_f, p_f \geq 0}
\left[ \alpha^2 {\Ga(-n_f + \alpha)^2 \ov \Ga(1+p_f - \alpha)^2 }
{\Ga(\q+2n_f+2i\sig_b+m_b-2\alpha) \ov \Ga(1\!-\!\q\!-\!2p_f\!-\!2i\sig_b+m_b + 2\alpha) }
\bar{q}_f^{n_f-\alpha} q_f^{p_f-\alpha}
\right]_\alpha \,,
\end{split}
\label{fint2}
\end{align}
where have defined
\be
p_f \equiv n_f -m_f \,.
\ee
We use the notation
\begin{align}
\left[ \cdots \right]_\alpha
&\equiv
{d \ov d\alpha} \left[ \cdots \right] |_{\alpha=0} \\
\left[ \cdots \right]_{\alpha\beta}
&\equiv
{d^2 \ov d\beta d\alpha} \left[ \cdots \right] |_{\alpha=0,\beta=0}
\end{align}
for convenience throughout this appendix.
Using the gamma function identity
\be
\Ga(z) \Ga(1-z) = {\pi \ov \sin \pi z}\,,
\ee
equation \eq{fint2} can be further reorganized into
\begin{align}
\begin{split}
\left[ (q_f \bar{q}_f)^{-\alpha}\left({\pi \alpha \ov \sin \pi \alpha} \right)^2
{\sin \pi(\q+2i\sig_b\!-\!m_b\!-\!2\alpha) \ov \pi}
f_{i \sig_b +{m_b \ov 2}+{\q \ov 2},\alpha} (\bar{q}_f)
f_{i \sig_b -{m_b \ov 2}+{\q \ov 2},\alpha} (q_f) \right]_\alpha \,,
\end{split}
\end{align}
where
\be
f_{z,\alpha} (q) \equiv \sum_{n \geq 0}
{\Ga (2n+2z-2\alpha) \ov \Ga (1+n-\alpha)^2} q^n
={\Ga(2z-2\alpha) \ov \Ga(1-\alpha)^2}
{}_3F_2 \left[ \,
\begin{matrix}
1, z-\alpha, z-\alpha+{1\ov 2}\\
1-\alpha, 1-\alpha
\end{matrix}
\, ; \, 4q
\right] \,.
\label{fza}
\ee

Plugging this result into the equation for $Z_{S^2}$, we get
\begin{align}
\begin{split}
Z_{S^2} = \Biggl[ \sum_{m_b} & e^{ -im_b \theta_b}
\int {d\sig_b  \ov 2\pi} e^{-4\pi i \xi_b \sig_b} 
{\Ga (-i\sig_b - {m_b \ov 2})^2 \ov \Ga (1+i\sig_b - {m_b \ov 2})^2}
{\sin \pi(\q+2i\sig_b\!-\!m_b\!-\!2\alpha) \ov \pi} \\
&(q_f \bar{q}_f)^{-\alpha}\left({\pi \alpha \ov \sin \pi \alpha} \right)^2
f_{i \sig_b +{m_b \ov 2}+{\q \ov 2},\alpha} (\bar{q}_f)
f_{i \sig_b -{m_b \ov 2}+{\q \ov 2},\alpha} (q_f)  \Biggr]_\alpha \,.
\end{split}
\label{ZS22}
\end{align}
Now we complete the integral with respect to $\sig_b$.
In the large $\xi_b$ limit, we may deform the contour of
integration for $\sig_b$ to the lower-half complex $\sig_b$
plane as was with the case of $\sig_f$.
The only poles of the integrand are at the loci
\be
-i\sig_b-{m_b \ov 2} = -n_b, \qquad n_b \in \ZZ_{\geq 0}
\quad \text{and} \quad n_b \geq m_b \,,
\ee
due to the gamma function $\Ga (-i\sig_b - {m_b \ov 2})$.
As was with the integral with respect to $\sig_f$, defining
\be
p_b \equiv n_b-m_b \,,
\ee
we can rewrite \eq{ZS22} as
\begin{align}
\begin{split}
Z_{S^2} = 
\Biggl[ &
\left( {\pi \alpha \ov \sin \pi \alpha}\right)^2
\left( {\pi \beta \ov \sin \pi \beta}\right)^2
(q_f \bar{q}_f)^{-\alpha}
(q_b \bar{q}_b)^{-\beta}
\left( {\sin \pi(\q-2\alpha-2\beta) \ov \pi }\right) \\
& \left( \sum_{n_f \geq 0} {f_{n_f-\beta+{\q \ov 2},\alpha}(\bar{q}_f) \ov \Ga(1+n_f -\beta)^2}
\bar{q}_b^{n_f} \right)
\left( \sum_{p_f \geq 0} {f_{p_f-\beta+{\q \ov 2},\alpha}({q}_f) \ov \Ga(1+p_f -\beta)^2}
{q}_b^{p_f} \right)
\Biggr]_{\alpha\beta} \,.
\end{split}
\label{ZS23}
\end{align}

Now $f_{n_f-\beta+{\q / 2},\alpha}(\bar{q}_f)$ and
$f_{p_f-\beta+{\q / 2},\alpha}({q}_f)$
both have singularities at $q_f={1/4}$
due to the singularity of the hypergeometric function ${}_3F_2$
at unit argument.
In fact
\begin{align}
\begin{split}
&f_{p_f -\beta+{\q \ov 2},\alpha} (q_f)  \\
&= {\Ga (-{1 \ov 2} +2p_f+\q-2\beta) \ov
{\sqrt{\pi} 2^{1-2p_f-\q+2\alpha+2\beta}}}
{}_3F_2 \left[ \,
\begin{matrix}
-\alpha, -\alpha, 0\\
\ha-p_f-{\q \ov 2}+\beta-\alpha, 1-p_f-{\q \ov 2}+\beta-\alpha
\end{matrix}
\, ; \, 1
\right] \\
& \quad \times(1-4q_f)^{1/2-2p_f-\q+2\beta}
+ \OO((1-4q_f)^{3/2-2p_f-\q+2\beta}) \\
&=
{\Ga (-{1 \ov 2} +2p_f+\q-2\beta) \ov {\sqrt{\pi} 2^{1-2p_f-\q+2\alpha+2\beta}}}
(1-4q_f)^{1/2-2p_f-\q+2\beta} + \OO((1-4q_f)^{3/2-2p_f-\q+2\beta}) \,,
\end{split}
\label{3f2}
\end{align}
where we have used results of \cite{Buhring}.\footnote{The
the argument $0$ appearing in the hypergeometric function
in equation \eq{3f2} forces it to have constant value $1$.}
In the field theory limit \eq{ftlimit} --- where we take
$\epsilon \ra 0$ --- the subleading terms of $(1-4q_f)$
can be ignored unless $p_f =0$.
When $p_f =0$, there is an order-one contribution we
cannot ignore in the small $(1-4q_f)$ limit and
the leading order expansion of
$f_{p_f-\beta+\q/2,\alpha}$ becomes
\begin{align}
\begin{split}
f_{-\beta+{\q \ov 2},\alpha} (q_f) 
= &{\Ga(\q-2\alpha-2\beta) \Ga({1 \ov 2} +2\beta -\q) \ov
 2^{\q+2\alpha-2\beta}\sqrt{\pi} \Ga(1-\q-2\alpha+2\beta)} \\
&+{\Ga (-{1 \ov 2} +\q-2\beta) \ov {\sqrt{\pi} 2^{1-\q+2\alpha+2\beta}}}
(1-4q_f)^{1/2-\q+2\beta} + \OO((1-4q_f)) \,,
\end{split}
\end{align}
Plugging in these expressions,
we see that $Z_{S^2}$ can indeed be written as an
expansion of $(\Lambda^4/u^2)$ in the limit \eq{ftlimit}
as
\be
 f_{p_f-\beta,\alpha}({q}_f) {q}_b^{p_f} \propto
\begin{cases}
\epsilon^{1-2\q} \sqrt{u} \left( {\Lambda^4 \ov u^2} \right)^{p_f}
+ \OO(\epsilon^{3-2\q}) & \text{when} ~p_f>0 \\
C_0+\epsilon^{1-2\q} \sqrt{u} \left( {\Lambda^4 \ov u^2} \right)^{p_f}
+ \OO(\epsilon^{2}) & \text{when}~ p_f=0 \\
\end{cases}
\ee
and likewise for its complex conjugate.

By evaluating \eq{ZS23} in the field theory limit,
we obtain at leading order in $\epsilon$
\be
Z_{S^2} = Z_L \ln \epsilon + Z_0 + \epsilon^{2-4\q} Z_{2-4\q} + \OO(\epsilon^2) \,.
\ee
$Z_L$ and $Z_0$ are both independent of $u$.
In fact, $Z_L$ is a function of $\q$
\be
Z_L =- {16 \cos (\pi \q) \Ga(\ha-\q)^2 \Ga(\q)^2 \ov 2^{2\q} \pi \Ga(1-\q)^2} \,,
\label{ZL}
\ee
while $Z_0$ is of the form
\be
Z_0 = f(\q) +g(\q) (\ln \Lambda + \ln \bar{\Lambda}) \,.
\label{Z0}
\ee
$Z_{2-4\q}$ is given by
\be
Z_{2-4\q} =-8\pi \sin ( \pi \q) Q\bar{Q}
-4 \cos (\pi \q) \left( Q \bar{Q}_D + \bar{Q} Q_D \right) \,.
\label{Z2}
\ee
$Q$ and $Q_D$ are defined as
\be
Q = (2u)^{-\q}\sqrt{u \ov \pi} \sum_{n \geq 0}
{\Ga (2n -\ha+\q) \ov \Ga(1+n)^2}
\left( {\Lambda^4 \ov 4u^2} \right)^n
\ee
and
\be
Q_D =(2u)^{-\q} \sqrt{u \ov \pi} \sum_{n \geq 0}
{\Ga (2n -\ha+\q) \ov \Ga(1+n)^2}
\left( \psi(n+1)-\psi (2n-\ha+\q ) + \ln (2u/\Lambda^2) \right)
\left( {\Lambda^4 \ov 4u^2} \right)^n \,.
\ee
Here $\psi$ is the digamma function
\be
\psi(z) = \Ga'(z)/\Ga(z) \,.
\ee

Therefore the quantum K\"ahler potential
of $\MM$ deformed by a small R-charge $\q$
is given by
\be
-\ln Z_{S^2} =
-\ln (Z_L \ln \epsilon+Z_0)
- {\epsilon^{2-4\q} \ov \ln \epsilon}
{Z_{2 -4\q} \ov Z_L} + 
\text{(sub-leading in $\epsilon$)}
\label{ZS2eps}
\ee
in the field theory limit.
We note that the first term can be
``gauged away" by a K\"ahler transformation
and does not affect the computation of the
K\"ahler metric on the moduli space of
the gauge theory.
The K\"ahler metric of the $SU(2)$
SYM moduli space can be obtained
from \eq{ZS2eps} by taking the small
$\q$ limit of
\be
-\p_u \p_{\bar{u}} \ln Z_{S^2} =
- {\epsilon^{2-4\q} \ov \ln \epsilon}
{\p_u \p_{\bar{u}} Z_{2 -4\q} \ov Z_L} + 
\text{(sub-leading in $\epsilon$)} \,.
\label{conc1}
\ee
We find that
\be
{\p_u \p_{\bar{u}} Z_{2 -4\q} \ov Z_L}
= {\q^2 \ov 4} \p_u \p_{\bar{u}}
\left( Q \bar{Q}_D + \bar{Q} Q_D \right) |_{\q=0} + 
\text{(sub-leading in $\q$)} \,.
\label{conc2}
\ee

Meanwhile, the Seiberg-Witten K\"ahler potential can be
written as
\be
K_{SW} (u,\Lambda)
= {1 \ov 2i} (\bar{a} a_D -a \bar{a}_D )
\ee
where
\begin{align}
{a \ov \Lambda} &= {\sqrt{2} \ov \pi}
\int_{-1}^1 {dx \sqrt{x-w} \ov \sqrt{x^2-1}} \\
{a_D \ov \Lambda} &= {\sqrt{2} \ov \pi}
\int_{1}^{w} {dx \sqrt{x-w} \ov \sqrt{x^2-1}} \,,
\end{align}
and $w \equiv u/\Lambda^2$.
$a$ and $a_D$ can be written in terms of hypergeometric functions \cite{Bilal:1995hc}
\begin{align}
{a \ov \Lambda} &= \sqrt{2 \left({w} +1 \right)} \,
F (-{1 \ov 2}, {1 \ov 2} ; 1; {2 \ov w+1})  \\
{a_D \ov \Lambda} &= i {(w -1 ) \ov 2} \,
F ({1 \ov 2}, {1 \ov 2} ; 2; {1-{w} \ov 2})  \,.
\end{align}

Using hypergeometric identities,
and the gamma function identity
\be
\Ga(2z) ={2^{2z-1} \ov \sqrt{\pi}} \Ga(z) \Ga(z+{1 \ov 2}),
\ee
one can show that
\begin{align}
\begin{split}
{a \ov \Lambda} &= \sqrt{2 \left({w} +1 \right)} \,
F (-{1 \ov 2}, {1 \ov 2} ; 1; {2 \ov w+1})  \\
&= \sqrt{2 w} ~
F (-{1 \ov 4}, {1 \ov 4} ; 1; {1 \ov w^2})  \\
&= -\sqrt{w \ov 2\pi} \sum_{n \geq 0}
{\Ga (2n -\ha) \ov \Ga(1+n)^2}
\left( {1 \ov 2w} \right)^{2n} = -{Q \ov \sqrt{2}\Lambda}
\bigg|_{\q =0}\,,
\end{split}
\label{a}
\end{align}
and that
\begin{align}
\begin{split}
{a_D \ov \Lambda} &= i {(w -1 ) \ov 2} \,
F ({1 \ov 2}, {1 \ov 2} ; 2;{1 -w \ov 2}) \\
&= i{(w-1) \ov 2}{(w+1) \ov 2}w^{-3/2}
F({3 \ov 4},{5\ov 4} ; 2; 1-{1 \ov w^2}) \\
&= {i \sqrt{w} \ov 4} {1 \ov (-{1 \ov 4})({1 \ov 4}) }
\lim_{c \ra 0} {1 \ov \Ga(c)} F(-{1 \ov 4} ; {1 \ov 4};c;1-{1 \ov w^2}) \\
&= {-4i\sqrt{w} \ov \Ga(-1/4)^2 \Ga(1/4)^2 } \\
&\times \sum_{n \geq 0}
{\Ga(n-{1 \ov 4}) \Ga(n+{1 \ov 4}) \ov \Ga(n+1)^2}
\left( 2\psi (n+1)-\psi(n-{1 \ov 4})-\psi(n+{1 \ov 4})+2\ln w \right) {1 \ov w^{2n}} \\
&=-{i \ov \sqrt{2} \pi} \sqrt{w \ov \pi}
\sum_{n \geq 0} {\Ga(2n-{1 \ov 2}) \ov \Ga(n+1)^2}
\left(\psi (n+1)-\psi(2n-{1 \ov 2})+\ln (2w) \right) {1 \ov (2w)^{2n}} \\
&= -{i \ov \sqrt{2} \pi} {Q_D \ov \Lambda}
\bigg|_{\q =0} \,.
\end{split}
\label{ad}
\end{align}

From \eq{conc1}, \eq{conc2}, \eq{a} and \eq{ad}
we find that
\be
-\lim_{\epsilon \ra 0} 
\left( - {\ln \epsilon \ov  {\epsilon^{2-4\q}}} \right)
 \p_u \p_{\bar{u}} \ln Z_{S^2}
= {\q^2 \pi \ov 2i}  \p_u \p_{\bar{u}} (\bar{a} a_D -a \bar{a}_D )
+ \text{(sub-leading in $\q$)}
\ee
when $\q$ is a small positive number.
The result \eq{mainresult} follows
accordingly.

\end{document}